# Analytical calculation of the charge spectrum generated by ionizing particles in Resistive Plate Chambers at low gas gain

**P. Fonte**[a,b]

[a] *LIP-Laboratório de Instrumentação e Física Experimental de Partículas,
Dep. de Física, Univ. de Coimbra, 3004-516 Coimbra, Portugal
E-mail*: fonte@coimbra.lip.pt

[b] *ISEC-Instituto Superior de Engenharia de Coimbra,
Rua Pedro Nunes - Quinta da Nora, 3030-199 Coimbra, Portugal*

ABSTRACT: In this article it is shown analytically that the charge spectrum generated by ionizing particles in Resistive Plate Chambers under Townsend avalanche conditions, that is, for sufficiently small avalanches not affected by space-charge and considering single-electron ionization clusters follows closely the statistical gamma distribution. This distribution describes well comparable simulation data taken from the literature, but seems to describe as well experimental data measured beyond these assumptions, rising some interpretation issues.

KEYWORDS: Particle tracking detectors (Gaseous detectors), Timing detectors.

## 1. Introduction

An analytical expression for the charge spectrum generated by ionizing particles in Resistive Plate Chambers (RPCs) would be of great theoretical interest for guiding detector design, verifying Monte-Carlo calculations (e.g. [1]) and analysing experimental results.

Expressions for the distributions arising from the successive ionization clusters deposited in the gas gap are known [1] but their sum, generating the whole distribution, was not derived.

In here it is shown that for avalanches containing a number of electrons large enough to justify a continuous description (around 100 electrons), single-electron ionization clusters and a Poissonian-distributed number of clusters the charge distribution follows closely a statistical gamma distribution. Comparisons with experimental data and Monte-Carlo calculations are presented and discussed.

## 2. Statement of the problem

Even if an RPC is irradiated with perfectly identical particles, the number of electrons generated by the avalanches will not be equal for each particle. This arises from avalanche gain fluctuations (process $\mathcal{A}$) and ionization statistics. The latter include contributions from cluster statistics* (process $\mathcal{B}$), the position of each cluster relative to the anode (process $\mathcal{C}$) and the variable number of clusters generated by each passing particle (process $\mathcal{D}$).

All contributions should be properly convoluted to yield the full distribution.

## 3. Avalanche multiplication

The stochastic evolution of a small avalanche progressing by a distance $z$ and generating the (random) charge $\mathcal{N}_e$ with probability distribution function (PDF) $P_\mathcal{A}(\mathcal{N}_e)$† is generally accepted as being given by Legler's avalanche theory. This has been calculated in a convenient way by Riegler [2]. For sufficiently large $\mathcal{N}_e$ (approx. $\mathcal{N}_e \geq 100$ - see, for instance, [5] Fig. 6) a continuous approximation is possible and the Laplace transform of $P_\mathcal{A}(\mathcal{N}_e)$, $M_\mathcal{A}(s)$, also called the moment-generating function (MGF), is given by

$$M_\mathcal{A}(s) = \frac{N_e(1-r)s + r}{N_e s + r}$$
$$r = 1 - \frac{\eta}{\alpha}, \ N_e = e^{\alpha^* z}$$
(1)

where $s$ is the complex frequency, $\alpha$ is the first Townsend coefficient, $\eta$ is the attachment coefficient and $\alpha^* = \alpha - \eta$ is the effective first Townsend coefficient. Laplace transform inversion yields

$$P_\mathcal{A}(\mathcal{N}_e) = (1-r)\delta(\mathcal{N}_e) + r\frac{r}{N_e} e^{-\mathcal{N}_e r/N_e}$$
(2)

This is actually a mixture of two distributions. A Dirac $\delta$ distribution at zero charge with weight $1-r$, corresponding to the probability that the avalanche will be extinguished owing to the electronegativity of the gas, and, with weight $r$, an exponential distribution with average value $N_e/r$. Therefore the average value of $P_\mathcal{A}(\mathcal{N}_e)$ is $E_\mathcal{A}(\mathcal{N}_e) = N_e$.

It should be noted that it is known that RPCs work typically in an avalanche regime that is strongly influenced by space-charge [3]-[7]. So, the present calculation is only valid for sufficiently small and independent avalanches, that is, for the low-charge region of the charge spectrum or at low gas gain. Any electrode-related effects are also neglected.

## 4. Ionization

### 4.1 Cluster statistics

The cluster statistics (process $\mathcal{B}$) can be handled analytically as shown in [2], yielding solutions that are expressed as integrals (z-transforms). Such expressions are quite opaque and equivalent results can be derived more easily by Monte-Carlo. As the results in [2] suggest that this is not a determinant feature in RPCs, we will neglect this process for the sake of reaching useful analytical expressions, leaving its treatment for a future work.

---

* The stochastic variation of the number of electron-ion pairs in each ionization cluster.
† Will denote in underscript the processes contributing the corresponding quantity.

**4.2 Cluster position**

The distribution of the amount of charge created from a single primary electron-ion pair anywhere in the gap (process $C$) is the randomization [8] of (2) on the parameter $z$ uniformly distributed over the gap width $d$:

$$P_{AC}(\mathcal{N}_e) = \int_0^d P_A(\mathcal{N}_e, z) \frac{1}{d} dz = (1-r)\delta(\mathcal{N}_e) + r \frac{e^{-\mathcal{N}_e r/G} - e^{-\mathcal{N}_e r}}{\mathcal{N}_e \ln(G)} \tag{3}$$

where $G = e^{\alpha^* d}$ is the maximum (cathode-to-anode) average gas gain. The corresponding MGF is

$$M_{AC}(s) = (1-r) + \frac{r}{\ln(G)} \ln\left(\frac{r+s}{r/G+s}\right) \tag{4}$$

In here we are slightly violating the conditions for a continuous approximation stated in the previous section because when the integral (3) approaches $z = 0$ very small avalanches will be considered. Therefore the present calculation is likely faulty for $\mathcal{N}_e \lesssim 100$.

The mean and variance for this distribution is

$$\begin{aligned} E_{AC}(\mathcal{N}_e) &= \frac{G-1}{\ln(G)} \simeq \frac{G}{\ln(G)} \\ E_{AC}(\mathcal{N}_e^2) - E_{AC}(\mathcal{N}_e)^2 &= \frac{G-1}{\ln(G)}\left(\frac{1}{r} - \frac{G-1}{\ln(G)}\right) \simeq \left(\frac{G}{\ln(G)}\right)^2 \end{aligned} \tag{5}$$

that is, for large gain and admitting that $r$ is not too small, the variance is the same as for an exponential distribution with the same average.

**4.3 Number of clusters**

The number $n$ of clusters (process $D$) is Poisson-distributed with PDF and probability-generating function, respectively,

$$\begin{aligned} P_D(n) &= \frac{e^{-\lambda d}(\lambda d)^n}{n!} \\ C_D(\zeta) &= e^{\lambda d(\zeta - 1)} \end{aligned} \tag{6}$$

where $\lambda$ is the average cluster density and therefore $\lambda d$ is the average number of clusters produced in the gas gap. For multigap RPCs with $N$ gaps it should be used $\lambda N$ instead of $\lambda$.

The distribution of the total generated charge will be the compounding [9] of (4) with (6):

$$M_{ACD}(s) = C_D(M_{AC}(s)) = \left(\frac{r+s}{r+Gs}\right)^m, \ m = \frac{\lambda r}{\alpha^*} \tag{7}$$

Analytic Laplace-inversion of (7) is only possible for integer $m$, yielding cumbersome but exact expressions for $P_{ACD}(\mathcal{N}_e)$ (omitted). A series expansion may be achieved by taking notice that[‡]

$$M_{ACD}(s) = \left(\frac{r+s}{r+Gs}\right)^m = \left(\frac{1}{G} + \frac{(G-1)}{G}\underbrace{\frac{r}{r+Gs}}_{M(s)}\right)^m$$

$$= \left(\frac{G-1}{G}\right)^m \left[M^m(s) + m\frac{M^{m-1}(s)}{G-1} + \frac{m(m-1)}{2!}\frac{M^{m-2}(s)}{(G-1)^2} + \right.$$

$$\left. + \frac{m(m-1)(m-2)}{3!}\frac{M^{m-3}(s)}{(G-1)^3} + \ldots \right]  \quad (8)$$

and that

$$P_k(\mathcal{N}_e) = \mathcal{L}^{-1}\left[M^k(s)\right] = \frac{e^{-\frac{\mathcal{N}_e}{G/r}}\mathcal{N}_e^{k-1}(G/r)^{-k}}{\Gamma(k)} \quad (9)$$

which for $k > 0$ is the statistical gamma distribution with shape parameter $k$ and scale parameter $G/r$. Note that the function is also defined for non-integer negative $k$ and that[§] $P_0(\mathcal{N}_e) = \delta(\mathcal{N}_e)$. Therefore

$$P_{ACD}(\mathcal{N}_e) = \left(\frac{G-1}{G}\right)^m \left[P_m(\mathcal{N}_e) + m\frac{P_{m-1}(\mathcal{N}_e)}{G-1} + \frac{m(m-1)}{2!}\frac{P_{m-2}(\mathcal{N}_e)}{(G-1)^2} + \right.$$

$$\left. + \frac{m(m-1)(m-2)}{3!}\frac{P_{m-3}(\mathcal{N}_e)}{(G-1)^3} + \ldots \right] \quad (10)$$

Note that for integer positive $m$ the series stops with the singular term $G^{-m}\delta(\mathcal{N}_e)$, representing the fundamental inefficiency $(1-\varepsilon)$ arising from either all avalanches being extinguished by the electronegativity of the gas or no cluster being produced. As

$$1 - \varepsilon = G^{-m} = e^{-\lambda dr} \quad (11)$$

and $\lambda dr$ is the average number of clusters that develop an avalanche, (11) is just the Poissonian probability that there will be no avalanches. This is smaller than the practical inefficiency as it assumes a near zero charge detection threshold.

If $G \gg 1$ (most likely the interesting practical case) one may consider a slightly inaccurate but handier approximation by noting that for large $G$

---

[‡] Generalized binomial expansion.
[§] $\mathcal{L}^{-1}\left[M^0(s)\right] = \mathcal{L}^{-1}(1) = \delta$.

$$\left(\frac{1}{G}+\frac{G-1}{G}\frac{r}{r+Gs}\right)^m \underset{G\gg 1}{\simeq} \left(\frac{r}{r+Gs}\right)^m \tag{12}$$

and therefore $P_{\mathcal{ACD}}(\mathcal{N}_e) \underset{G\gg 1}{\simeq} P_m(\mathcal{N}_e)$. Keeping the singular term as well, which represents a fundamental physical feature, and normalizing to unity we form the approximation – our main result –

$$P_{\mathcal{ACD}}(\mathcal{N}_e) \underset{G\gg 1}{\simeq} \tilde{P}_{\mathcal{ACD}}(\mathcal{N}_e) = G^{-m}\delta(\mathcal{N}_e) + (1-G^{-m})P_m(\mathcal{N}_e) \tag{13}$$

The mean and variance are

$$\begin{aligned}
E_{\mathcal{ACD}}(\mathcal{N}_e) &= (1-G^{-m})m\frac{G}{r} \\
&= (1-G^{-m})\lambda d\frac{G}{\ln(G)} \\
E_{\mathcal{ACD}}(\mathcal{N}_e^2) - E_{\mathcal{ACD}}(\mathcal{N}_e)^2 &= (1-G^{-m})(1+mG^{-m})m\frac{G^2}{r^2} \\
&= (1-G^{-m})(1+mG^{-m})\frac{\lambda d\ln(G)}{r}\left(\frac{G}{\ln(G)}\right)^2
\end{aligned} \tag{14}$$

One recognizes that the average generated charge is the efficiency times the average number of clusters from (6) times the average gain from (5) (for large $G$). If the inefficiency is small $(G^{-m}\ll 1)$ the relative standard deviation is just $1/\sqrt{m}$. A multigap RPC, where $N$ identical gaps contribute simultaneously to the signal, is equivalent to increasing $\lambda$ to $\lambda N$ in a single gap, therefore reducing the relative standard deviation by a factor $1/\sqrt{N}$. Comparison with (5) allows to conclude that process $\mathcal{D}$ changes the variance of the distribution over the single-cluster case by a factor approximately $\lambda d \ln(G)/r$, which is likely much larger than unity for most practical situations.

The accuracy of the approximation (13) is illustrated via a numerical example in Fig. 1 by comparison between the exact $P_{\mathcal{ACD}}(\mathcal{N}_e)$ available for integer $m \geq 1$ and $\tilde{P}_{\mathcal{ACD}}(\mathcal{N}_e)$.

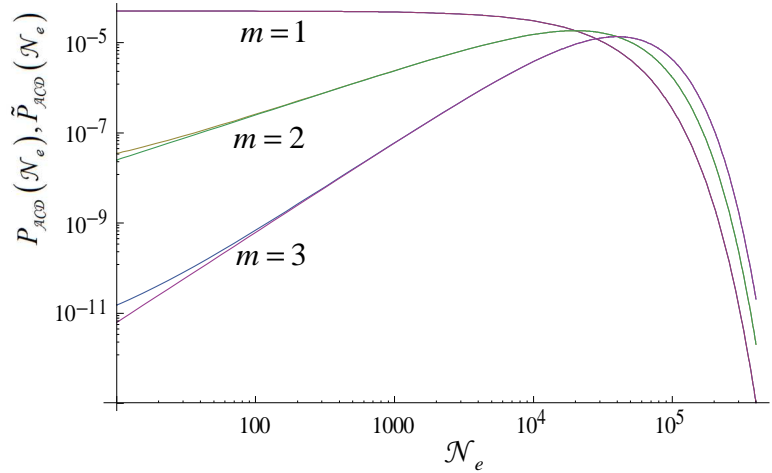

Fig. 1 – Comparison between the exact $P_{ACD}(\mathcal{N}_e)$ (upper curve for each $m$) and $\tilde{P}_{ACD}(\mathcal{N}_e)$ (lower curve for each $m$) for $G = 10^4$, $r = 0.5$, $m = \{1, 2, 3\}$, suggesting that the approximation (13) is sound for sufficiently large $\mathcal{N}_e$.

Note that for $\mathcal{N}_e \ll G/r$, $\tilde{P}_{ACD}(\mathcal{N}_e)$ follows a power law: $\tilde{P}_{ACD}(\mathcal{N}_e) \sim \mathcal{N}_e^{m-1}$. Therefore the parameter $m = r\lambda/\alpha^*$, essentially the ratio between the cluster density and the effective first Townsend coefficient, controls qualitatively the behaviour of $\tilde{P}_{ACD}(\mathcal{N}_e)$ for small $\mathcal{N}_e$: if $m < 1$ the function diverges at origin and it is monotonically decreasing, while for $m > 1$ it is null at the origin. This has been already noted in [1].

## 5. Comparison with simulations and data

As the comparison between experimental data and a Monte-Carlo simulation equivalent to the present paper's was already performed [1], in here we will just compare with said Monte-Carlo. In Fig. 2 it is represented $\tilde{P}_{ACD}(\mathcal{N}_e)$ for the conditions corresponding to Figure 12 of [1]. The similarity is quite striking, even if in the simulation there seems to be more events at the larger charges. This may be a statistical effect owed to the small number of events in such region.

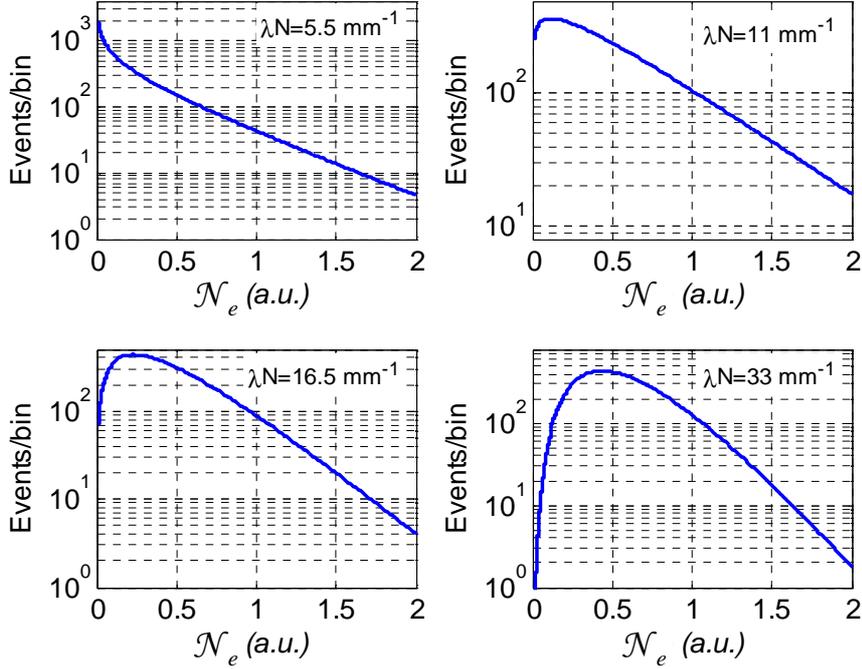

Fig. 2 - Representation of $\tilde{P}_{ACD}(\mathcal{N}_e)$ for conditions similar to those in Figure 12 of [1]. For all cases $\alpha^* = 9\,mm^{-1}, r = 1$. The scale parameter $G/r$ and the vertical scale were adjusted to obtain a similar appearance as the units cannot be directly compared.

It has been observed that, as expected, for thin gas gaps the charge distribution follows a power law with $m < 1$ close to the origin, deviating from this law for larger, space-charge influenced, avalanches [4]. However, it seems that the gamma distribution also adjusts multigap RPCs' distributions. In Fig. 3 it is presented the adjustment of $\tilde{P}_{ACD}(\mathcal{N}_e)$ to experimental data collected from the devices described in [10]. These are 5-gaps RPCs, with 2960V applied to each 0.35 mm wide gap filled with a mixture of $C_2H_2F_4/SF_6$ 90/10. The particles were almost vertical cosmic rays. A similar exercise is made in Fig. 4 for a 4-gaps RPC in a pion beam [11].

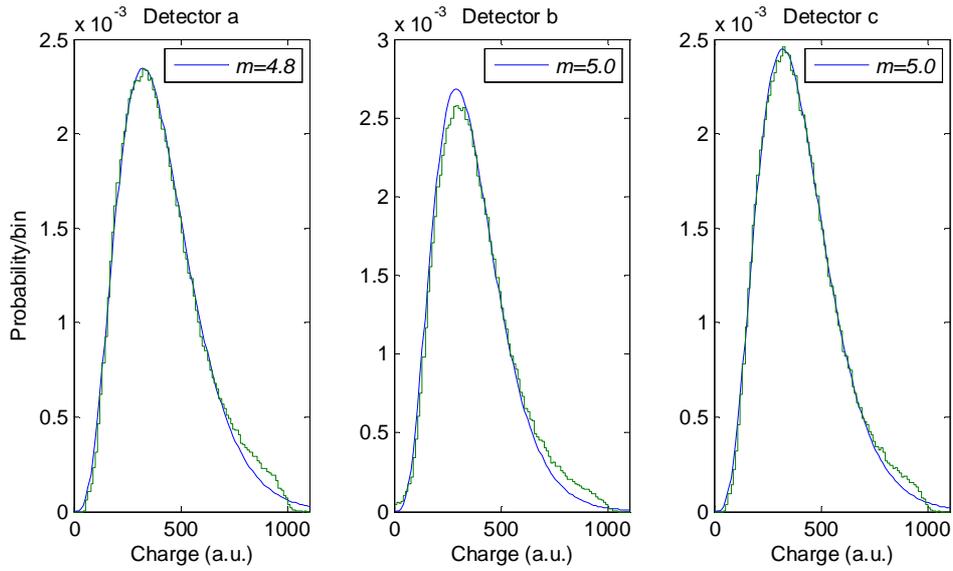

Fig. 3 – Adjustment of $\tilde{P}_{ACD}(\mathcal{N}_e)$ to experimental data collected from [10] showing a remarkable agreement.

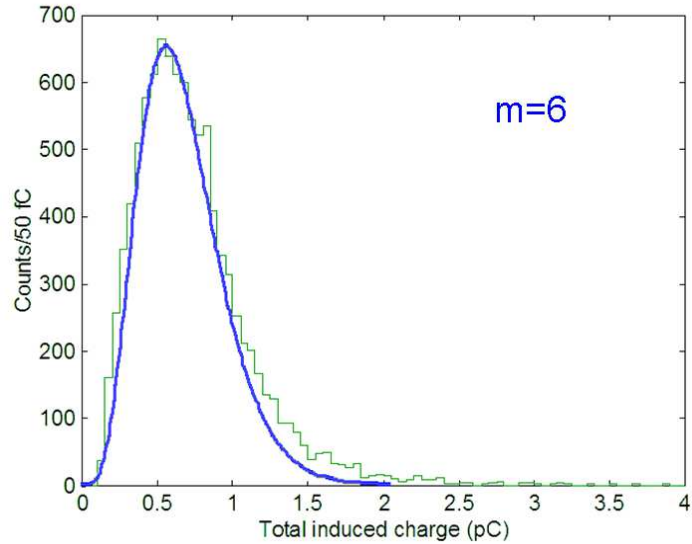

Fig. 4 - Comparison of $\tilde{P}_{ACD}(\mathcal{N}_e)$ to experimental data presented in [11] showing a quite good agreement.

## 6. Discussion

The apparent excellent empirical agreement between the experimental data and $\tilde{P}_{ACD}(\mathcal{N}_e) \approx P_m(\mathcal{N}_e)$ (statistical gamma distribution), suggesting that the experimental distribution arises from the mechanisms calculated here, arises serious problems of interpretation.

For a typical applied electric field in 0.3 mm gaps of around 90 kV/cm the effective first Townsend coefficient will be close to 100 mm$^{-1}$ for the base gas ($C_2H_2F_4$) typically used [12].

– 8 –

(Addition of a small amount of $SF_6$ changes the applied field by just a percentage.) In these conditions and taking the experimental value $m = \lambda Nr / \alpha^* \approx 5$, the number of gas gaps $N \approx 5$ and $r \leq 1$ we conclude that the cluster density $\lambda$ should be at least close to 100 mm$^{-1}$, which seems to be a complete impossibility for minimum ionizing particles. On this line of thought one would be led to hypothesise the efficient emission into the gas gap of highly ionizing particles, a process that hasn't been identified so far.

An alternative, in line with the current understanding of the workings of RPCs, would be that the $\sim 1/\mathcal{N}_e$ single electron distribution (3) would be really not so owing to the strong influence of the space charge effect ([3]-[7]), which strongly modifies the single-gap distribution (as actually measured [4]), reducing its variance. In this case the apparent agreement between the multigap RPC's charge distribution and the gamma distribution would be a statistical accident or property (maybe somewhat similar to the Central Limit Theorem) arising from the convolution of $N$ single-gap space-charge modified distributions to create the $N$-gaps distribution, as it is actually demonstrated in [11].

## 7. Conclusion

The charge distribution generated by RPCs in absence of a space-charge effect and neglecting cluster statistics follows closely a statistical gamma distribution. This compares very well with Monte-Carlo calculations and with experimental data.

However the meaning of this agreement is open to discussion, as it seems to apply as well to space-charge dominated situations and the measured variance is too small for comfortable physical interpretation.

In any case, it seems that the gamma distribution may be a convenient representation of the charge distribution in some RPC configurations, either from theoretical or empirical grounds.

## Acknowledgment

This work was financed by the Portuguese Government through FCT-Foundation for Science and Technology and FEDER/COMPETE under the contract CERN/FP/123605/2011.## References